# Can scientists and their institutions become their own open access publishers?

Translator and Editorial Consultant, Granada, Spain
(Note 1)

Correspondence:
Karen Shashok
C./ Compositor Ruiz Aznar 12, 2-A
18008 Granada, Spain
Tel: +34 958 812305, email: kshashok@kshashok.com
ORCID: 0000-0002-2506-1390

**Abstract**

This article offers a personal perspective on the current state of academic publishing, and posits that the scientific community is beset with journals that contribute little valuable knowledge, overload the community's capacity for high-quality peer review, and waste resources. Open access publishing can offer solutions that benefit researchers and other information users, as well as institutions and funders, but commercial journal publishers have influenced open access policies and practices in ways that favor their economic interests over those of other stakeholders in knowledge creation and sharing. One way to free research from constraints on access is the diamond route of open access publishing, in which institutions and funders that produce new knowledge reclaim responsibility for publication via institutional journals or other open platforms. I argue that research journals (especially those published for profit) may no longer be fit for purpose, and hope that readers will consider whether the time has come to put responsibility for publishing back into the hands of researchers and their institutions. The potential advantages and challenges involved in a shift away from for-profit journals in favor of institutional open access publishing are explored.

**¿Pueden los científicos y sus instituciones convertirse en editoriales de acceso abierto por sí mismos?**

**Resumen**

Este artículo ofrece una perspectiva personal a propósito del estado actual de la edición académica, y propone que la comunidad científica se encuentra lastrada por las muchas revistas que contribuyen pocos conocimientos de valor, sobrecargan la capacidad común de proporcionar una revisión experta de calidad, y desperdician los recursos. La edición en acceso abierto puede ofrecer soluciones que benefician a los investigadores y otros usuarios de la información, además de las instituciones y los patrocinadores, pero las editoriales comerciales de revistas científicas han influido en las políticas y prácticas del acceso abierto mediante vías que favorecen sus intereses económicos por encima de los intereses de otras partes interesadas en la creación y diseminación de conocimientos. Una manera de liberar a la investigación de las restricciones al acceso es la vía diamante de edición en acceso abierto, en la cual las instituciones y los patrocinadores que producen los nuevos conocimientos reclaman la responsabilidad de la edición a través de revistas institucionales u otras plataformas abiertas. Propongo que las revistas de investigación (sobre todo aquellas que son editadas como productos comerciales) ya no cumplen con su finalidad original, y espero que los lectores se planteen si es oportuno o no devolver a los investigadores y sus instituciones la responsabilidad de la edición y diseminación. Se exploran las ventajas potenciales así como los retos relacionados con el abandono progresivo de las revistas comerciales a favor de la edición institucional en acceso abierto.

**Palabras clave:** acceso abierto, calidad de la investigación, centros de investigación, diamante, dorado, edición académica, editor, ética, financiación, instituciones, investigadores, revisión por expertos, revistas, partes interesadas, verde

**************
## Note 1. Why does open access matter to me?

As a science-technical-medical translator and authors' editor [29] I often need to learn technical terminology and concepts quickly. But I cannot afford publishers' paywalls for useful-looking articles. Some publishers' online journal platforms make it difficult to find the corresponding author's email address to request a copy – or do not provide one at all. Some authors never get the final pdf of their own articles [6,17] or are afraid to share them even in response to individual requests for research or teaching purposes. Some researchers have told me that the publisher's terms and conditions forbid person-to-person sharing. Some who have complied with take-down notices have told me they were afraid that noncompliance would place them on a blacklist and make it harder for them to publish again in the same journal or other journals owned by the same publisher.

    For me as a user of information, the current system is an obstacle that sometimes prevents me from doing my work as well as I could. For researchers more generally, restrictions on sharing undermine their efforts to contribute to knowledge, and restrictions on access limit their efforts to build on current knowledge.

/**************

**Can scientists and their institutions become their own open access publishers?**

Is academic publishing in peer-reviewed journals fit for purpose? It depends on who you ask. The main commercial beneficiaries of the current system, heavily dominated by for-profit journals and tools for research quantification and evaluation, are keen defenders of their economic model. In contrast, pioneers in open access (OA) to research argue that digital technologies are available to make publication, access and evaluation better than they are now [33]. Better how? In ways that benefit all stakeholders in research production, communication and application, except for commercial enterprises that have grown and hugely benefited from the outsourcing and privatization, by institutions that produce research, of both publication and research evaluation.

In the now over-long and painful process of transforming research publication to take full advantage of digital information management technologies, many major initiatives have included commercial publishers as sources of input and guidance. Instead of truly innovating, most big publishers have used these opportunities, together with their own political and economic lobbying initiatives, to add relatively minor tweaks to the publishing system in ways that ultimately favor their interests over the interests of other stakeholders. These publishers are unlikely to take the lead in reforms that would require them to renounce their main sources of profit: pressure to publish and research quality evaluation systems based on journal ranking mainly with the impact factor (IF).

Are journals produced by commercial publishers providing high-quality services to researchers? Again, it depends on who you ask. Editor and publisher errors apparently caused by ignorance, incompetence, carelessness and lack of suitable quality controls are on the rise. Although publishers will usually claim that errors are rare and affect only a very small proportion of all articles, the increasing incidence of complaints from other stakeholders is a sign that the traditional system of journal publication is no longer as trustworthy as it once was. Meadows and Wulf described seven areas where journal publishers' failure to understand researchers' needs sometimes leads to considerable inconvenience for manuscript authors, and to errors in the publication process [31].

The risk of publication process errors is increased by large commercial publishers' decentralization of publishing tasks, coupled with a lack of strong systems of editorial oversight. At commercial academic journals the editor may not feel directly responsible for anything that happens to a manuscript outside the peer review and selection process. But little is known about editors' formal duties and responsibilities because their contracts are not made public and so cannot be the subject of rigorous research. Editorial and administrative staff – often outsourced, and often (under)paid by volume of work – may not feel accountable to anyone other than their contact person, and may not feel any particular loyalty to the journal brand. As a result, external manuscript editors, copyeditors, typesetters and coders may not be aware of the importance of error prevention at every step in the publication process. Publishing imprints are usually owned by multinational corporations whose central mission is removed from academic publishing – an environment where editorial quality may not be a high priority for market managers and decision-makers. As long as the publishing arm continues to bring in profits from selling subscriptions and article processing charges (APCs), corporate managers probably feel that their academic knowledge market is insensitive to issues with declining journal quality.

**Issues with editorial quality control**

Is there is a shortage of good peer reviewers? Here too, it depends on who you ask. A small minority of very prestigious journals can attract reviewers easily, whereas editors at other journals struggle to find competent reviewers who provide useful feedback in a timely manner. For many journals, reviewers have little incentive to donate time and hard intellectual work since public recognition for their help may be meager, and employers still do not consider good peer review a merit-worthy contribution to science. Perhaps part of the reason for the scarcity of reviewing capacity is because new journals are launched to occupy emerging market niches faster than sufficient numbers of researchers can learn reviewing skills.

In my experience of more than 30 years working with researchers, the general quality of peer review has declined, and serious errors during production and publication have become more frequent. Authors need to work harder than ever to discover useful feedback in reviewers' reports, and to reconcile conflicting demands by different reviewers. Revising the manuscript is made especially challenging when authors are caught between mutually incompatible requests, especially if the editor provides no advice other than "please revise your manuscript accordingly".

Competition for publication in prestigious journals worked as a driver of quality while research funding was generous, but cutbacks have made competition so vicious that publication misconduct and other forms of cheating have proliferated. Editors and reviewers are poorly equipped to detect intentional fraud and are generally reluctant to accept responsibility for this. The publication of unreliable work would not be so problematic if journals were responsive to post-publication alerts about serious problems, but mechanisms for correcting the record are used inconsistently, and in many cases the editor or publisher fails to take appropriate action. Südhof recently observed that,

> "...there is little accountability for journals and reviewers. If a journal repeatedly publishes papers that draw untenable conclusions, eventually the authors of the papers may be blamed, but editors and reviewers who are arguably responsible for gross negligence are not held responsible. There are insufficient checks and balances in the publishing system; when high-ranked journals repeatedly publish papers that are later considered unreliable or even retracted, the journals seem to face no consequences—their premier status remains untouched" [48].

**Publisher errors and interference**

Readers are reporting more errors in published articles and making more requests for corrections, but publishers seem unable to stay up to date in correcting the scientific record. A study of editorial quality and key errors that eluded detection before publication found that the time-to-correction (the time between the date of article publication and the date of the correction) has increase since 1993 at three prestigious journals: Science, Nature and PNAS. The time-to-correction was longer for severe mistakes affecting the reliability of the data or conclusions than for minor errors. There was no correlation between the frequencies of published corrections and the IF across a sample of 16 journals that included both large general science journals and major specialty journals [27].

The Retraction Watch blog lists entries for retractions motivated by publisher errors [43]. The reasons for these retractions are "accidental" duplicate (or even triplicate) publication, duplicate manuscript submittal by authors detected after publication, lack of correspondence between the online ahead of print and final printed versions of the same paper, publication in the wrong journal, failure to publish conflict of interest statements, fake or biased peer review detected after publication [44], publication of a rejected manuscript or of the wrong version of a manuscript, undeclared editorial conflict of interest, failure to detect and reject obviously poor or fraudulent research, and various administrative errors. These are examples of the growing number of cases in which editors and publishers have not handled key tasks competently.

Typesetting errors are on the rise, often because of incompetent file conversion and weak quality assurance during journal production. Mathematicians and chemists, in particular, are becoming disgusted with the time they waste undoing the damage caused by editorial or production staff [34-36], and errors are also a source of concern for modelers, statisticians and other researchers who use highly formulaic, standardized nomenclature and symbols to communicate their work.

Examples of publication delays because of conflicts over the content abound. One particularly painful case involved editorial interference with an article about researcher self-publishing, in a Taylor & Francis journal [21,37,38]. More recently Tennant described his frustrations with peer review, production and publication at a Wiley journal [49]. In 2016 a colleague and I (unbeknownst to each other at the time) spent ridiculous amounts of time dealing with inefficient editorial practices at a major journal (Note 2).

\*\*\*\*\*\*\*\*\*\*\*\*\*\*
**Note 2. Dear Editor, We read with interest…**

A colleague and I traded notes after we published letters, several months apart, in response to different items that appeared in a large general research journal with a double-digit impact factor. We had each assumed that our experience was uncharacteristically complicated, but were surprised to discover later how much of our frustration with the process was caused by the same things. One of us had to deal with intrusive copyediting that changed the meaning, introduced errors and required prompt, repeated action by the author to correct. We both encountered a lack of coordination between editorial staff members we corresponded with, and spent time resolving an issue over the journal's absurd policy not to publish non-institutional affiliations (we are both self-employed). Editorial changes were made after acceptance (the journal reserves this right), yet further changes were made even after the journal had sent us its final version for approval. This journal does not send proofs of letters, so authors have no opportunity to see and approve what the journal ultimately publishes. We also found editorial staff to be unfamiliar with the journal's own access policies – a situation that led one editor to recommend that one of us add a comment to our letter online, apparently without realizing that 1) letters at that time were not open access, and 2) commenting would require purchase of online access in order to enable the online commenting function. (This solution was of course unacceptable; fortunately the journal found a work-around.) Although each of our letters links back to the article we commented on, there is no link forward from the main items to our letters. In fact, there is no indication in the main items online that the journal has published comments about them.

The time we wasted dealing with the publishing process, all for letters containing just 88 and 193 words, led each of us to vow never to submit anything to this journal again. What we experienced was not the level of service to authors and readers one expects from a top international journal.

/**************

Publishers often claim to be committed to the widest possible dissemination of articles, but most commercial publishers use closely controlled terms and conditions of dissemination that prevent many users from consulting published research. One result is the disturbing trend toward the "enclosure of scholarly infrastructure" [5]. The merger of Springer and Nature Publishing Group, the acquisition of Mendeley and SSRN by Elsevier [28], and Elsevier's PURE tool [19] are examples of corporate strategies to consolidate the academic journal oligopoly, and to amalgamate and control as many links as possible in the chain of knowledge communication and sharing.

**Gold open access: an unsustainable option?**

The conversion by commercial publishers to OA financed through APCs instead of subscriptions (flipping) may not be sustainable in the long term. Moreover, flipping would perpetuate the flow of public or philanthropic funds toward private profit, and perpetuate the current inequities in the affordability of OA for many researchers, institutions and countries.

Several analyses have noted the drawbacks of hybrid OA options offered by commercial publishers as a model for gold OA. Harnad has always considered this approach "fool's gold" because it makes little economic sense for research funders to pay APCs as long as green OA via repositories is an option [20]. Björk has reasoned that if the hybrid OA solution provides "a vehicle for the successful transformation of leading subscription publishers to OA publishers", the result may be that "the major subscription publishers could end up dominating the OA market, charging academia roughly the same amount of money for their services as before, and their profit levels would remain the same as before" [6]. The Pay It Forward Study, which investigated the sustainability of APC-based OA, concluded that "[f]or the most research-intensive North American research institutions, the total cost to publish in a fully article processing charge-funded journal market will exceed current library journal budgets" [54]. In a 2014 report prepared at the request of the European Commission, Archimbault et al. noted that, "[t]he current model of back end toll access is simply unsustainable because of the gross social inefficiency and ineffectiveness" – a situation "that taxpayers the world over should not tolerate". Regarding institutional mandates for gold OA, these authors concluded that favoring a shift from reader-pays access to author-pays access could transform access from an issue of inaccessibility into an issue of inequality. In their words, "[n]either inaccessibility nor growing inequality are acceptable considering that universalism is one of the core values of scientific research" [2].

Fuchs and Sandoval renamed gold OA "corporate OA", noting that APCs favor commercial publishers' interests above all other stakeholders. These authors pointed out that discussions about gold OA tend to focus on the purported need for APCs to cover publishers' costs, while overlooking the fact that most OA journals do not charge APCs, or charge fees much lower than those levied by hybrid OA journals. In addition, they noted that journal coverage by Web of Science seems clearly skewed in favor of hybrid journals with a gold OA option, whereas coverage of OA journals with low or no publication fees is disproportionately sparse and unrepresentative of the fact that low- or no-APC OA journals far outnumber hybrid OA journals [16].

A report by Solomon et al. [47] focused on the flip to OA in 15 different journals – 10 that depend on APCs and 5 that do not. These authors provide examples of economic and management strategies, evidence of changes that were or were not successful, and their own analysis of the strengths and weaknesses of different flipping strategies. Their analysis looked at the APC conversion process through the lens of economic sustainability for subscription-based publishers, the underlying assumptions being that journals are and will remain the main package for delivering new knowledge, and must remain profitable or at least not loss-making. Brembs explained the gold OA situation for researchers, funders and information users bluntly:

> "What determines how much we are going to pay for an OA article as long as we have for-profit publishers in there is not going to be what their costs are; it's going to be how much can they can charge for it and still survive and actually make a profit that is better than their competitors" [8].

Librarians and information scientists were among the first to warn policymakers of the excessive costs of journal subscriptions and APCs, and to propose more cost-effective alternatives. Although this article does not cover their contributions to debates about OA and institutional publishing in depth, it is important to acknowledge their role as researchers and stakeholders in the publishing system. Table 1 provides a few sources of additional information.

**Table 1. Librarians as sources of open access research and expertise**

| Source | Available at | Notes |
| --- | --- | --- |
| Library Publishing Coalition (LPC) | http://www.librarypublishing.org/about-us . Accessed 7 Sept 2016 | "Based on core library values, and building on the traditional skills of librarians, [library publishing] is distinguished from other publishing fields by a preference for Open Access dissemination as well as a |

| | | willingness to embrace informal and experimental forms of scholarly communication and to challenge the status quo." |
|---|---|---|
| The LPC Bibliography | http://www.librarypublishing.org/resources/biblio | Lists articles published from 2013 to the present about the key contributions of institutional librarians and information scientists to institution-led publishing |
| University of California Libraries. Pay it Forward Team. Pay it Forward. Investigating a Sustainable Model of Open Access Article Processing Charges for Large North American Research Institutions | http://icis.ucdavis.edu/wp-content/uploads/2016/07/UC-Pay-It-Forward-Final-Report.rev_.7.18.16.pdf . Accessed 9 Aug 2016 | A large-scale study by the University of California, Davis, and the California Digital Library on behalf of the University of California Libraries, with collaboration from libraries at Harvard University, Ohio State University and the University of British Columbia. The finding shed "new light on the financial viability of the article processing charge business model to create open access at a much larger scale". |
| University of Cambridge Office of Scholarly Communication. Unlocking Research Blog | https://unlockingresearch.blog.lib.cam.ac.uk/?page_id=2 . Accessed 13 Sept 2016 | Maintained by the Office of Scholarly Communication based in the University of Cambridge Library and the University Research Office, the threads in this blog cover scholarly communication, open research, open access, research data management, and library and training matters. |
| Walters T. The Future Role of Publishing Services in University Libraries. In Libraries and the Academy, Vol. 8, No. 4 (2008), pp. 425–454. Baltimore: The Johns Hopkins University Press. | https://www.press.jhu.edu/journals/portal_libraries_and_the_academy/portal_pre_print/articles/12.4walters.pdf . Accessed 7 Sept 2016 | From the abstract: "The study participants comprised university library directors, library managers responsible for publishing services, and library association personnel and consultants involved in publishing. Many participants saw collaborating with multiple libraries and other stakeholder organizations to establish publishing cooperatives as essential." |

An adverse but perhaps predictable result of the hybrid journal model based on APCs is the OA gold rush of predatory journal publishers that compete with legitimate journals for authors' funds. As in any uncontrolled, unregulated competition for something of value, those who act in good faith are deceived and elbowed out by cheaters, eager to exploit an opportunity for easy gains before an authority steps in to provide order. Commercial journal publishers were quick to exploit APCs, arguing that they were entitled to this source of revenue to replace income lost from cancelled subscriptions. Unfortunately, setting up a decoy online journal to attract APCs is simple, and disguising the website as a legitimate journal requires only a little more effort by scammers. When researchers mistake a fake journal for a legitimate one, they not only waste their time and money, but may find they cannot withdraw and resubmit their work elsewhere because they cannot recall their copyright transfer. In the USA the Federal Trade Commission recently brought legal action against one publisher because it has "deceiv[ed] readers about reviewing practices, publication fees, and the nature of its editorial boards" [30].

The demarcation between predatory and legitimate journals, however, is sometimes blurred by a tendency to label legitimate journals as "predatory" on the basis of assumptions about how or where they operate. Moreover, established commercial publishers also engage in predatory-like behavior when they provide poor service to authors, readers and institutions despite payment of an APC [3,13,24,53].

**Repositories and online networks: a necessary transitional step?**

Online open repositories for preprints and postprints are available to facilitate access to research (see for example [4]). So far two main types of repository have appeared. Larger, discipline-centered depots include arXiv, bioRxiv, chemRxiv [1], SSRN and SocArXiv, to be launched as a replacement for SSRN [9]. In addition, institutional repositories are being created by individual organizations, funders and consortia (see for example [57,58]). Repositories increasingly provide tools for users to review, annotate and evaluate items, and these functions make repositories a good way to meet the needs of most stakeholders.

As one of the foundations of green OA, repositories are a partial solution to truly open access but seem to create as many problems as they solve. If an institutional repository interface is hard to learn and use, researchers will not be inclined to spend undue time uploading each preprint or postprint. The repository version may not be appropriately labeled or may not be linked to the final (non-OA) version. For information users on a tight deadline, it can be frustrating to find a desired item on an open repository, only to have to search further to identify the journal and locate the table of contents in order to obtain full bibliographic details for citation. A further drawback of repositories created to support green OA is that within the broader publication ecosystem, they are ultimately a work-around that consumes funder's resources to accommodate commercial publishers' embargoes on access to the version of record.

Social academic networks like Academia.edu and ResearchGate are often easier and faster for researchers to use than discipline-related or institutional repositories. But if these networks operate as businesses, their long-term availability is uncertain because they are subject to economic forces rather than the needs of the scientific community [6]. In other words, commercial networks that cease to be profitable may disappear with no warning. Or they may simply be acquired and terminated by competitors.

**Diamond open access: institutions as publishers**

Fuchs and Sandoval [16] worried that publication in commercial journals uses research funding in an unsustainable manner by feeding a system that has taken control of research communication away from scientists and the academy. As a way to reclaim the academic commons, they proposed diamond OA publishing. The diamond route is defined by University of Groningen Library as a model that

> "differs from gold open access in that the costs of editing, peer review, online publication, hosting, etc., are borne by an institution, fund or collaborative arrangement. Societies, universities and other noncommercial institutions make an infrastructure available and most of the professional work is done by academics in their roles as editors or peer reviewers" [55].

As explained by Fuchs and Sandoval, diamond OA publishing "can realise the true essence of academia as a communication system that produces and communicates academic knowledge as a commons in an open process". They further note that a shift to this route will require "public funding, policies that base evaluations and research grants on the diamond model and a system of rewards for scholars who act as editors, editorial board members or reviewers for such publications". In summary, they recommend bringing all research dissemination activities back under the purview of research funders, particularly public institutions, and eliminating commercial intermediaries from the system.

Open access journals published with support from research institutions [32] are another alternative with close parallels to diamond OA, and institutional journals could of course coexist with other institutional publishing platforms. In either package the diamond OA model returns control of publishing to the academic and scholarly community, where it was as recently as the mid-20th century before the boom in commercial journal publishing. The researcher-driven diamond route can in fact be seen not as a type of OA, and not as an ideological goal supported by members of some movement, but simply as a return to the way scientists used to publish, free from third-party service provider-imposed constraints to access and the economic indenture that have resulted from outsourcing key responsibilities for scholarly communication to commercial intermediaries.

If most research is funded by taxpayers and not-for-profit institutions, why are commercial enterprises being allowed to control publication and access according to market interests that work against the widest possible dissemination of research results? Bringing research publication and evaluation back under the purview of public institutions and other not-for-profit funders could have a number of advantages. For example, institution-based open access research publishing has the potential to:

- reduce the overall costs of dissemination and access,
- increase transparency and accountability,
- save funds now being spent on the race to publish as much as possible,
- increase the speed and efficiency of dissemination and access,
- reduce global inequities in the creation of new knowledge, and

- forestall the trend toward enclosure of publication and evaluation by commercial companies.

Fewer, better journals [22] and a shift toward research publishing in large, open, institutional or discipline-focused platforms could make knowledge sharing more efficient and cost-effective. Institutional diamond OA publishing could greatly reduce the volume of information in the research literature if institutions focused on publishing only their best work and refrained from disseminating preliminary results or least publishable units. As a result, the scientific community's currently overstretched capacity for rigorous peer review is more likely to be sufficient for all new contributions. Moreover, it would become easier for users to stay up to date with new publications.

Staff experts in methodology, statistics, ethics, writing and reporting would be key sources of expertise in ensuring quality control and enhancing the institution's reputation for careful, rigorous work. Publication support provided by in-house colleagues is likely, I believe, to be more effective in upholding quality than the current circuitous route to publication through intermediaries. Currently, gaps in reviewers' and editors' expertise allow flawed work to get through to publication, and mechanisms for correcting the scientific record often fail because of publishers' and institutions' lack of motivation or conflicting priorities. With institution-managed publishing, the marker of rigor and quality would be the institution, not the journal brand. In addition, institutional OA platforms could be used to make valuable knowledge available in other formats [58] such as methodology notes, datasets, replications, bibliographies and theses. Routes to research dissemination independently of journal publishers are now being actively explored [24,40),41]

**Challenges and changes**

Resources now made available for gold OA publication and journal access should be redirected toward publication under the aegis of institutions and funders, with no need for commercial intermediaries that, I argue here, are no longer serving science or society well. Could institutions (e.g. universities, research centers, libraries, scientific societies) offer researchers a better way to publish? At his Open and Shut blog Poynder observed that, "many are concluding that it is time for the research community to wean itself off for-profit publishers". He noted, however, that efforts by the academic community "to recover its ownership of scholarly communication, and in the process regain control of costs" undoubtedly face significant challenges [42]. Quantifying the total cost of publication is complex because publishers make different deals with their institutional clients (the terms of which are often subject to confidentiality clauses), and institutions themselves do not always have accurate records of their publication costs [39]. Estimating the cost of creating and running institutional journals and platforms for research dissemination is likewise difficult. Table 2 provides some current sources of information on the costs of open access publishing.

**Table 2. Blogs, reports and research on the costs of open access**

| Source | Available at | Notes |
|---|---|---|
| Bergstrom TC. Ted Bergstrom's Journal Pricing Page | http://econ.ucsb.edu/~tedb/Journals/jpricing.html . Accessed 10 Sept 2016 | Main site for economist Ted Bergstrom's work on journal pricing |
| Bergstrom TC, Courant P, McAfee RP. Big Deal Contract Project | http://econ.ucsb.edu/~tedb/Journals/BundleContracts.html . Accessed 10 Sept 2016 | This group of economists, "[a]s citizens of the academic community, are interested in helping librarians to understand the dynamic economic problem that they face and aiding them in negotiating effectively with large publishers. We plan to release a collection of information and analyses that will serve this purpose." |
| Bergstrom TC, Courant P, McAfee RP. Journal cost-effectiveness 3013 | http://www.journalprices.com/ . Accessed 10 Sept 2016 | A compendium of data on journal prices, citations, and number of articles published, with estimates of value per dollar for each of about 7,000 journals. Searchable by journal title, publisher and ISSN. |
| Bergstrom TC, Courant PN, McAfee RP, Williams MA (2014) Evaluating big deal journal bundles. Proc Nat Acad Sci 111(26):9425–9430 | http://www.pnas.org/content/111/26/9425.full.pdf . Accessed 10 Sept 2016 | |
| Brembs B. How Gold Open Access May Make Things Worse. bjoern.brembs.blog The blog of neurobiologist Björn Brembs | http://bjoern.brembs.net/2016/04/how-gold-open-access-may-make-things-worse/ . Posted 7 Apr 2016. Accessed 5 Aug 2016 | |
| Brembs B. What Interacting With Publishers Felt Like For This Open Access Proponent. bjoern.brembs.blog The blog of | http://bjoern.brembs.net/2016/08/what-interacting-with-publishers-felt-like-for-this-open-access-proponent/. Posted 1 Aug 2016. Accessed 3 Aug 2016. | |
| neurobiologist Björn Brembs. | | |
| Crawford W. Cites & Insights: Crawford at large | http://citesandinsights.info/ . Accessed 4 Aug 2016 | Crawford, a semi-retired library writer, editor, speaker, researcher and systems analyst, writes an ongoing series of thorough, data-rich critical analyses of developments in open access publishing. |
| Crawford W. Gold Open Access Journals 2011-2015 (2016) Cites & Insights: Crawford at Large 16(5) June | http://citesandinsights.info/ci v16i5.pdf . Accessed 4 Aug 2016 | An overview of global costs of OA journal publishing categorized in three knowledge areas (human and social sciences, hard sciences-technology-economics-mathematics, and biomedicine) and covering article volumes of journals, fees and revenue, publisher category, country of publication, regions and APCs, viability, gray OA (delisting) and the Directory of Open Access Journals. |
| Crawford W. Gold Open Access Journals 2011-2015 (2016) Cites & Insights Books, Livermore, California | http://www.lulu.com/us/en/sh op/walt-crawford/gold-open-access-journals-2011-2015/paperback/product-22758867.html . Free version available at http://waltcrawford.name/goa j1115.pdf . Accessed 10 Sept 2016 | "This book reports on a comprehensive analysis of serious open access journals as of December 31, 2015: nearly 11,000 journals in the Directory of Open Access Journals. For 10,324 of the journals, the study includes whether or not there's an article processing charge (APC), how much it is, and the number of articles in each year 2011 through 2015. The state of serious gold OA is described in terms of article volume, fees and revenue, subject segments, regions, type of publisher and other aspects." |
| Harnad S. Publishing Costs. Open Access Archivangelism | http://openaccess.eprints.org/i ndex.php?/categories/18- | |
| | Publishing-Costs . 14 April 2016. Accessed 7 Sept 2016 | |
| Pinfield S, Salter J, Bath PA (2016) The "Total Cost of Publication" in a Hybrid Open-Access Environment: Institutional Approaches to Funding Journal Article-Processing Charges in Combination With Subscriptions. Journal of the Association for Information Science and Technology 67(7):1751–1766 | http://onlinelibrary.wiley.com/doi/10.1002/asi.23446/full . Accessed 9 Sept 2016 | |
| Shamash K. Article processing charges (APCs) and subscriptions. Monitoring open access costs (Report). Joint Information Systems Committee (JISC). | https://www.jisc.ac.uk/reports/apcs-and-subscriptions Published 27 June 2016. Accessed 7 Sept 2016 | At UK universities hybrid journals accounted for 80% of APC expenditure in 2014-2015. APCs account for <15% of publication costs (subscriptions + APCs) but this proportion is expected to increase. |
| Solomon D, Björk B-C (2016) Article processing charges for open access publication – the situation for research intensive universities in the USA and Canada. PeerJ 4:e2264; DOI 10.7717/peerj.2264 | https://peerj.com/articles/2264/ . Accessed 10 Sept 2016 | |
| Tennant JP, Waldner F, Jacques DC, Masuzzo P, Collister LB, Hartgerink CHJ (2016) The academic, economic and societal | http://f1000research.com/articles/5-632/v1 . Published online 2016 Jun 9. doi: 10.12688/f1000research.8460.2. PMCID: PMC4837983. Accessed 2 Sept 2016 | This review contains a useful summary of information from studies about the economics of OA publishing. |
| impacts of Open Access: an evidence-based review. Version 2. F1000Res 5: 632 | | |
| West JD, Bergstrom T, Bergstrom CT (2014) Cost Effectiveness of Open Access Publications. Economic Inquiry 52(4):1315–1321 | http://onlinelibrary.wiley.com/doi/10.1111/ecin.12117/abstract . Accessed 10 Sept 2016 | |

If stakeholders worked together, economic and human resources already available might be more than sufficient to move research dissemination to institutional platforms. A recent overview of the impact of OA concluded that "[f]or libraries, universities, governments, and research institutions, one major benefit of lowering the cost of knowledge is a budget that allows them to spend their resources more wisely" [50]. If institutions invest in building OA repositories or other platforms, it makes sense for institutions, their researchers and information users, rather than commercial corporations, to be the main beneficiaries of these investments. As explained by Haspelmath, a major goal of institutional diamond OA publishing would be to "create new prestigious labels that belong to us, the scientists, to give us freedom of publication. Publication labels are actually even more important to us than our campuses, and even more intrinsically connected to our careers and to our research environment" [22].

Another challenge would be to reform the way research productivity is evaluated and rewarded. The interdependence of commercial journal publishing and IF-based research evaluation was noted by Tracz in his talk "Life after the death of science journals" at the 2016 Researcher to Reader conference [52]. Responding to a comment about researchers' lack of motivation to use publisher-independent platforms rather than journals, Tracz observed that it is "unbelievably hard to stop the IF", and that the only way this could happen is if journals were brought to an end. Tracz saw the role of funders is a key factor in this change, noting that if funders decide they want researchers to publish on platforms, "there will simply not be a place for journals". Innovative publishing platforms were described in a recent evidence-based review of the economic and societal impacts of OA [50] that considered the perspectives of a number of stakeholders including policymakers, publishers, research funders, governments, learned societies, librarians, and academic communities.

Institutional publishing should be associated with professional prestige and merit, so the journal IF would need to be replaced with different approaches to evaluation that emphasize researchers and the quality of their work. By halting the race to publish and replacing it with incentives that favor scientific rigor and transparency, institutions could foster the publication of fewer, better articles. This measure could remove the perverse incentives that lead to the publication of unreliable work [11,23] and prevent the damage to an institution's reputation that results from the publication of unsound work. In addition, institutions and funders could place greater value on evidence of altruism and openness as the main drivers of knowledge sharing. As Brembs noted in 2013,

> "by overcoming journal rank and replacing it with a scientific reputation system as part of an institution-based publishing service for scholarly literature, software and data, we could collectively free more than US$9b every year for science and innovation. By further delaying publishing reform, we not only keep wasting tax-payer money, we also continue to reward salesmen who may possibly also be great scientists (if we are lucky) and to punish excellent scientists who are not extraordinary marketers" [7].

Publication in commercial research journals has become an outmoded, often untransparent and unaccountable, wasteful and dysfunctional system to communicate research. These journals are no longer fit-for-purpose because they do not meet the needs of researchers and society on a global level for rapid, efficient access to and exchange of information. Funders and institutions around the world have begun to resist pressures from commercial journal publishers to perpetuate costly subscription and APC agreements that allow research dissemination and access only under the publisher's terms [10,18,45,51,56]. The current burden of unneeded journals, many with ineffectual quality control, is in itself a waste of resources. Yet publishers force institutions to subscribe to unwanted and unused journals as part of their big deals – an abusive negotiating strategy that librarians have been denouncing for years. Journals and articles that contain unreliable information waste the funding used to produce and report the research, waste the resources publishers use to process submissions, waste resources used by indexers and aggregators to cover the many thousands of articles that are never read or cited, and waste readers' time in the process of literature searching and review. So many unreliable and unused publications are being added to the literature that the overall trustworthiness of published research is being compromised [13].

Will stakeholders in research put a stop to this cascade of waste? The League of European Research Universities has supported plans by the 2016 Dutch European Union Presidency to reduce the flow of research funds to commercial publishers [26]. Some funders have begun to re-examine how they can better support the cost-effective sharing of research results, and some national governments and international agencies, most notably the European Commission [12,14,15,58], have begun to rethink how to make the best use of OA. Although the EC Guidelines on Open Access to Scientific Publications and Research Data in Horizon 2020 [15] appear to consider journal publication as the preferred mode of dissemination, participants at the EC-sponsored workshop "Alternative Open Access Publishing Models" in October 2015 emphasized the need to move away from traditional publishers and APC-funded OA [46]. An important step back from reliance on third-party publishing has been taken by the Wellcome Foundation, which recently launched its own OA journal [25].

Commercial academic publishers justify their economic models by repeating that somebody has to pay for their publishing services to keep their system sustainable. Actually, somebody already *is* paying for and otherwise subsidizing their system, and that somebody is *us*: researchers, unpaid editors and reviewers, funders, libraries, producers and users of new knowledge, taxpayers, citizens. It is time for stakeholders to come together and try to make research publishing an open enterprise that everyone, both within and outside academia, can benefit from.

The year 2016 marked a move from denial to acceptance of the notion that commercial journal publishing may be dying as the main medium for disseminating new knowledge. Stakeholders should now look ahead and plan for a transition to new, open platforms. In a future without for-profit journals, the survivors and descendants of the old system could use digital, financial and institutional resources to move ahead and flourish by rebuilding a more efficient, more transparent and more globally equitable system of research communication and sharing.

**Acknowledgments**

I thank Joy Burrough-Boenisch for helpful feedback on part of the manuscript. My appreciation to all open access pioneers and innovative thinkers who have discussed in more depth and detail many of the issues raised in this article. No funding was provided for writing this article.

**References**

1. American Chemical Society (2016) American Chemical Society announces intention to establish "ChemRxiv" preprint server to promote early research sharing. Press Release 10 Aug 2016. https://www.acs.org/content/acs/en/pressroom/newsreleases/2016/august/acs-announces-intention-to-establish-chemrxiv-preprint-server-to-promote-early-research-sharing.html . Accessed 11 Aug 2016

2. Archambault E, Amyot D, Deschamps P, Nicol A, Provencher F, Rebout L, Roberge G (2014) Proportion of Open Access Papers Published in Peer-Reviewed Journals at the European and World Levels—1996–2013. European Commission, Brussels. http://science-metrix.com/files/science-metrix/publications/d_1.8_sm_ec_dg-rtd_proportion_oa_1996-2013_v11p.pdf . Accessed 22 Aug 2016

3. Bauer H (2013) Decadent science: Does fake differ from genuine? If so, how? Scepticism about Science and Medicine blog. https://scimedskeptic.wordpress.com/2013/04/24/decadent-science-does-fake-differ-from-genuine-if-so-how/ Posted 24 April 2013. Accessed 23 Oct 2016

4. Berg JM, Bhalla N, Bourne PE, Chalfie M, Drubin DG, Fraser JS, Greider CW, Hendricks M, Jones C, Kiley R, King S, Kirschner MW, Krumholz HM, Lehmann R, Leptin M, Pulverer B, Rosenzweig B, Spiro JE, Stebbins M, Strasser C, Swaminathan S, Turner P, Vale RD, VijayRaghavan K, Wolberger C (2016) Preprints for the life sciences. Science 352(6288): 899-901. doi: 10.1126/science.aaf9133. http://science.sciencemag.org/content/352/6288/899.full . Accessed 18 Aug 2016

5. Bilder G. The enclosure of scholarly infrastructure (2015) OpenCon 2015. 14–16 November 2015, Brussels. https://www.youtube.com/watch?v=oWPZkZ180Ho . Accessed 18 Aug 2016

6. Björk B-C (2016) The open access movement at a crossroad: Are the big publishers and academic social media taking over? Learned Publishing 29(2): 131–134. http://onlinelibrary.wiley.com/doi/10.1002/leap.1021/full?scrollTo=references . Uncorrected proof provided by author 23 Aug 2016

7. Brembs B (2013) By replacing journal rank with an institution-based reputation system, the looming crisis in science can be averted. London School of Economics and Political Science. Impact of Social Sciences. http://blogs.lse.ac.uk/impactofsocialsciences/2013/07/30/solutions-to-the-looming-crisis-in-science/ . Posted 30 July 2013. Accessed 17 Aug 2016

8. Brembs B. Making open default (2015) OpenCon 2015. 14–16 November 2015, Brussels. https://www.youtube.com/watch?v=pMxKW8g0SZc . 15 Nov 2015. Accessed 18 Aug 2016

9. Cohen PN. Announcing the development of SocArXiv, an open social science archive. https://socopen.org/2016/07/09/announcing-the-development-of-socarxiv-an-open-social-science-archive/ . July 2016. Accessed 25 Aug 2016

10. Earney L (2016) Jisc Collections and Elsevier agreement: questions and answers. Jisc blog. https://www.jisc.ac.uk/blog/jisc-collections-and-elsevier-agreement-questions-and-answers-28-nov-2016 . Posted 28 Nov 2016. Accessed 3 Jan 2017

11. Edwards MA, Roy S (2016) Academic Research in the 21st Century: Maintaining Scientific Integrity in a Climate of Perverse Incentives and Hypercompetition. Environmental Engineering Science. doi: 10.1089/ees.2016.0223. http://online.liebertpub.com/doi/full/10.1089/ees.2016.0223 . Online Ahead of Print 22 Sept 2016. Accessed 26 Sept 2016

12. Enserink M (2016) In dramatic statement, European leaders call for 'immediate' open access to all scientific papers by 2020. http://www.sciencemag.org/news/2016/05/dramatic-statement-european-leaders-call-immediate-open-access-all-scientific-papers . Posted 27 May 2016. Accessed 16 Aug 2016

13. Eriksson S, Helgesson G (2016) The false academy: predatory publishing in science and bioethics. Med Health Care and Philos. http://link.springer.com/article/10.1007%2Fs11019-016-9740-3 . Published online 7 October 2016. Accessed 19 Oct 2016

14. European Commission (2016) Press Release. Scientific data: open access to research results will boost Europe's innovation capacity. Brussels, 17 July 2016. http://europa.eu/rapid/press-release_IP-12-790_en.htm?locale=en . Accessed 5 Aug 2016

15. European Commission, Directorate-General for Research & Innovation (2016) H2020 Programme. Guidelines on Open Access to Scientific Publications and Research Data in Horizon 2020. Version 3.1, 25 August 2016. http://ec.europa.eu/research/participants/data/ref/h2020/grants_manual/hi/oa_pilot/h2020-hi-oa-pilot-guide_en.pdf . Accessed 25 Aug 2016

16. Fuchs C, Sandoval M (2013) The Diamond Model of Open Access Publishing: Why Policy Makers, Scholars, Universities, Libraries, Labour Unions and the Publishing World Need to Take Non-Commercial, Non-Profit Open Access Serious. tripleC 13(2):428-443. ISSN 1726-670X. http://www.triple-c.at or http://www.triple-c.at/index.php/tripleC/article/view/502 . Accessed 23 Aug 2016

17. Gardner CC, Gardner GJ (2015) Bypassing Interlibrary Loan Via Twitter: An Exploration of #icanhazpdf Requests. ACRL March 25-28 http://www.ala.org/acrl/sites/ala.org.acrl/files/content/conferences/confsandpreconfs/2015/Gardner.pdf . Accessed 17 Aug 2016

18. Gowers T (2016) Time for Elsexit? Gowers's Weblog. https://gowers.wordpress.com/2016/11/29/time-for-elsexit/ . Posted 2 Nov 2016. Accessed 30 Nov 2016

19. Harnad S (2015) Elsevier's PURE: self-interest and exploitation. Open Access Archivangelism. http://openaccess.eprints.org/index.php?/archives/1164-.html . Posted 12 Nov 2015. Accessed 17 Aug 2016

20. Harnad S (2016) Publishing Costs. Open Access Archivangelism. http://openaccess.eprints.org/index.php?/categories/18-Publishing-Costs . 14 April 2016. Accessed 7 Sept 2016

21. Harvie D, Lightfoot G, Lilley S, Weir K (2014) Publisher, be damned! From price gouging to the open road. Prometheus 31(3): 229–239. http://www.tandfonline.com/doi/full/10.1080/08109028.2014.891710 . Accessed 20 Aug 2016

22. Haspelmath M (2015) How to switch quickly to diamond open access: The best journals are free for authors and readers. Free Science Blog. http://www.frank-m-richter.de/freescienceblog/2015/10/28/how-to-switch-quickly-to-diamond-open-access-the-best-journals-are-free-for-authors-and-readers/ . Posted 28 Oct 2015. Accessed 23 Aug 2016

23. Higginson AD, Munafò MR (2016) Current Incentives for Scientists Lead to Underpowered Studies with Erroneous Conclusions. PLoS Biology http://dx.doi.org/10.1371/journal.pbio.2000995 and http://journals.plos.org/plosbiology/article?id=10.1371/journal.pbio.2000995 . Published 10 Nov 2016. Accessed 12 Nov 2016

24. Lagoze C, Edwards P, Sandvig C, Plantin J-C (2015) Should I stay or should I go? Alternative infrastructures in scholarly publishing. International Journal of Communication 9: 1052–1071. http://ijoc.org/index.php/ijoc/article/view/2929 . Accessed 23 Oct 2016

25. Lawrence R (2016) Wellcome Open Research: the start of a new journey. F1000 Research. http://blog.f1000research.com/2016/11/15/wellcome-open-research-the-start-of-a-new-journey/ . Posted 15 November 2016. Accessed 19 Nov 2016

26. League of European Research Universities (LERU) (2015) Christmas is over. Research funding should go to research, not publishers! LERU Statement for the 2016 Dutch EU Presidency. http://www.leru.org/index.php/public/extra/signtheLERUstatement/ . Dated 12 Oct 2015. Accessed 25 Aug 2016

27. Margalida A, Colomer MÀ (2016) Improving the peer-review process and editorial quality: key errors escaping the review and editorial process in top scientific journals. PeerJ 4:e1670. https://doi.org/10.7717/peerj.1670 or https://peerj.com/articles/1670/ . Published 9 Feb 2016. Accessed 25 Aug 2016

28. Masnick M (2016) Just As Open Competitor To Elsevier's SSRN Launches, SSRN Accused Of Copyright Crackdown. Techdirt. https://www.techdirt.com/articles/20160718/02211935003/just-as-open-competitor-to-elseviers-ssrn-launches-ssrn-accused-copyright-crackdown.shtml . Posted 18 July 2016. Accessed 21 July 2016

29. Matarese V (2016) Editing Research. Information Today, Medford NJ. http://books.infotoday.com/books/Editing-Research.shtml

30. McCook A (2016) U.S. government agency sues publisher, charging it with deceiving researchers. Retraction Watch. http://retractionwatch.com/2016/08/26/u-s-government-group-sues-publisher-charging-it-with-deceiving-researchers/#more-43653 . Posted 26 Aug 2016. Accessed 26 Aug 2016

31. Meadows A, Wulf K (2016) Seven Things Every Scholarly Publisher Should Know about Researchers. Scholarly Kitchen. https://scholarlykitchen.sspnet.org/2016/08/30/seven-things-every-scholarly-publisher-should-know-about-researchers/ . Posted 30 Aug 2016. Accessed 10 Sept 2016

32. Momen H (2014) Institutional journals as an alternative model for open access. Memórias do Instituto Oswaldo Cruz 109(7): 847–848. doi: 10.1590/0074-0276140334 http://www.ncbi.nlm.nih.gov/pmc/articles/PMC4296487/ . Accessed 20 Aug 2016

33. Moody G (2016) Open access: All human knowledge is there—so why can't everybody access it? Techdirt http://arstechnica.com/science/2016/06/what-is-open-access-free-sharing-of-all-human-knowledge/ . Posted 17 June 2016. Accessed 17 Aug 2016

34. Murray-Rust P (2013) Why should we continue to pay typesetters/publishers lots of money to process (and even destroy) science? And a puzzle for you. https://blogs.ch.cam.ac.uk/pmr/2013/02/21/why-should-we-continue-to-pay-typesetterspublishers-lots-of-money-to-process-and-even-destroy-science-and-a-puzzle-for-you/ . Posted 21 Feb 2013. Accessed 25 Aug 2016

35. Murray-Rust P (2014) Publishers' typesetting destroys science: They are all as bad as each other. Can you spot the error? https://blogs.ch.cam.ac.uk/pmr/2014/12/13/publishers-typesetting-destroys-science-they-are-all-as-bad-as-each-other-can-you-spot-the-error/ . Posted 13 Dec 2014. Accessed 25 Aug 2016.

36. Murray-Rust P (2015) This month's typographical horror: Researchers PAY typesetters to corrupt information. https://blogs.ch.cam.ac.uk/pmr/2015/01/19/this-months-typographical-horror-researchers-pay-typesetters-to-corrupt-information/ . Posted 19 Jan 2015. Accessed 25 Aug 2016

37. Perakakis P (2014) Academic self-publishing: a not-so-distant-future. Open Scholar. http://www.openscholar.org.uk/academic-self-publishing-a-not-so-distant-future/ . Posted 19 June 2014. Accessed 20 Aug 2016

38. Perakakis P, Taylor M (2013) Academic self-publishing: a not-so-distant future. Prometheus 31(3): 257-263. http://dx.doi.org/10.1080/08109028.2014.891712 . Accessed 20 Aug 2016

39. Pinfield S, Salter J, Bath PA (2015) The "Total Cost of Publication" in a Hybrid Open-Access Environment: Institutional Approaches to Funding Journal Article-Processing Charges in Combination With Subscriptions. Journal of the Association for Information Science and Technology 67(7): 1751–1766 http://onlinelibrary.wiley.com/doi/10.1002/asi.23446/full . Accessed 9 Sept 2016

40. Poss R, Altmeyer S, Thompson M, Jeller R (2014) Academia 2.0: Removing the publisher middle-man while retaining impact. Conference paper. 1st ACM SIGPLAN Workshop on Reproducible Research Methodologies and New Publication Models in Computer Engineering (TRUST'14), Edinburgh, United Kingdom. doi: 10.1145/2618137.2618139. https://www.researchgate.net/publication/262846836_Academia_20_Removing_the_publisher_middle-man_while_retaining_impact . Accessed 23 Oct 2016

41. Poss R, Altmeyer S, Thompson M, Jeller R. Aca 2.0 Q&A (2014) Usage scenarios and incentive systems for a distributed academic publication model. https://arxiv.org/pdf/1404.7753.pdf . 3 June 2014. Accessed 23 Oct 2016

42. Poynder R (2016) SocArXiv debuts, as SSRN acquisition comes under scrutiny. Open and Shut. http://poynder.blogspot.com.es/2016/07/socarxiv-debuts-as-ssrn-acquisition.html#more . Posted 19 July 2016. Accessed 21 July 2016

43. Retraction Watch (2016) http://retractionwatch.com/category/by-reason-for-retraction/publisher-error/. Accessed 03 Jan 2017

44. Retraction Watch (2016) Archive for the 'self peer review' category. http://retractionwatch.com/category/by-reason-for-retraction/self-peer-review/ . Accessed 7 Sept 2016

45. Schiermeier Q, Rodríguez Mega E (2016) Scientists in Germany, Peru and Taiwan to lose access to Elsevier journals. Nature 23 December 2016. http://www.nature.com/news/scientists-in-germany-peru-and-taiwan-to-lose-access-to-elsevier-journals-1.21223 . Posted 23 December 2016, Corrected 3 Jan 2017. Accessed 3 Jan 2017

46. Smith A (2015) Alternative Open Access Publishing Models: Exploring New Territories in Scholarly Communication. European Commission, Directorate General for Communications Networks, Content and Technology, Directorate C Excellence in Science, Brussels. https://ec.europa.eu/futurium/en/system/files/ged/oa_report.pdf . Accessed 20 Aug 2016

47. Solomon D, Laakso JM, Björk B-C (authors) (2016) In Suber P (editor). Converting Scholarly Journals to Open Access: A Review of Approaches and Experiences. Harvard University DASH Repository. http://nrs.harvard.edu/urn-3:HUL.InstRepos:27803834 . Accessed 5 Aug 2016

48. Südhof TC (2016) Truth in Science Publishing: A Personal Perspective. PLoS Biol 14(8):e1002547. doi:10.1371/journal.pbio.1002547. http://journals.plos.org/plosbiology/article?id=10.1371/journal.pbio.1002547 . Published 26 Aug 2016. Accessed 3 Sept 2016

49. Tennant J (2016) Why I will never publish with Wiley again. Green tea and velociraptors. https://fossilsandshit.com/2016/08/05/why-i-will-never-publish-with-wiley-again/ . Posted 5 Aug 2016. Accessed 25 Aug 2016

50. Tennant JP, Waldner F, Jacques DC, Masuzzo P, Collister LB, Hartgerink CHJ (2016) The academic, economic and societal impacts of Open Access: an evidence-based review. Version 2. F1000Res 5:632. doi: 10.12688/f1000research.8460.2. PMCID: PMC4837983. http://f1000research.com/articles/5-632/v1 . Published online 2016 Jun 9. Accessed 2 Sept 2016

51. Tiedonhinta.fi (2016) The cost of scientific publications must not get out of hand. http://tiedonhinta.fi/en/english/ . Accessed 30 Nov 2016

52. Tracz V (2016) Life after the death of science journals. Researcher to Reader Conference, 15–16 February 2016, London. http://river-valley.zeeba.tv/life-after-the-death-of-science-journals/ . Accessed 17 Aug 2016

53. Truth F (2012) Pay big to publish fast. Academic journal rackets. J Critical Educ Pol Studies 10(2): 54-105. http://www.jceps.com/wp-content/uploads/PDFs/10-2-02.pdf . Accessed 23 Oct 2016

54. University of California Libraries, Pay it Forward Team (2016) Pay it Forward. Investigating a Sustainable Model of Open Access Article Processing Charges for Large North American Research Institutions. http://icis.ucdavis.edu/wp-content/uploads/2016/07/UC-Pay-It-Forward-Final-Report.rev_.7.18.16.pdf . June 30, 2016, Revised July 18, 2016. Accessed 9 Aug 2016

55. University of Groningen Library (2016) Types of open access. http://www.rug.nl/bibliotheek/services/openaccess/vormen?lang=en . Last modified 22 July 2016. Accessed 22 Aug 2016

56. Vogel G (2016) Thousands of German researchers set to lose access to Elsevier journals. Science Insider. http://www.sciencemag.org/news/2016/12/thousands-german-researchers-set-lose-access-elsevier-journals#disqus_thread . Posted 22 Dec 2016. Accessed 23 Dec 2016

57. Wellcome Trust (2016) Wellcome Open Research. http://wellcomeopenresearch.org/ . Accessed 5 Aug 2016

58. Williams LS, Pope KH, Lucero BL (2014) Institutional repositories provide an ideal medium for scholars to move beyond the journal article. The Impact Blog. The London School of Economics and Political Science. http://blogs.lse.ac.uk/impactofsocialsciences/2014/03/12/institutional-repositories-move-beyond-the-journal-article/ . Posted 12 March 2014. Accessed 22 July 2016

**About the author**

The views expressed in this article distill my professional experience as a science-technical-medical translator and authors' editor, an author and peer reviewer, and an observer of research publishing processes since the mid-1980s. My thinking is informed by conversations with researchers from around the world as they struggle to reconcile conflicting career goals. They wish to publish the best research they can, and hope that it will be seen and used as widely as possible. But journals create inconsistent and sometimes unfair rules in the publication game, and pressures from research evaluators to publish more, publish faster, and publish in higher-impact journals can force researchers to opt for certain types of journals over more open or more trustworthy publishing options. The experiences of researchers I work with, and my own experiences as an author and user of research literature, are evidence that third-party control of publication and access by commercial service providers is not serving science well.